\documentclass[prd,notitlepage,nofootinbib,superscriptadress,12pt]{revtex4-1}

\usepackage{amsmath, amssymb}
\usepackage[]{graphicx}
\usepackage[caption=false]{subfig}
\usepackage{hyperref}
\usepackage{dsfont}
\usepackage{xcolor}
\usepackage{amsmath,amssymb,amsfonts, bm,bbm,slashed, subdepth}
\usepackage[normalem]{ulem}
\usepackage{hyperref}
\usepackage{cleveref}
\usepackage{enumerate}
\usepackage{setspace}
\usepackage{booktabs, tabularx}
\usepackage{units}

\usepackage[paper=a4paper,
  includefoot, 
  marginparwidth=0.0mm, 
  marginparsep=1.3mm, 
  margin=20mm, 
  includemp]{geometry}

\newcommand{\be}{\begin{equation}}
\newcommand{\ee}{\end{equation}}
\newcommand{\ba}{\begin{array}}
\newcommand{\ea}{\end{array}}
\newcommand{\bea}{\begin{eqnarray}}
\newcommand{\eea}{\end{eqnarray}}
\newcommand{\balg}{\begin{align}}
\newcommand{\ealg}{\end{align}}
\newcommand{\bit}{\begin{itemize}}
\newcommand{\eit}{\end{itemize}}
\newcommand{\trm}[1]{\textrm{#1}}

\newcommand{\Mpc}{\trm{\Mpc}}
\newcommand{\yr}{\trm{\yr}}
\newcommand{\eV}{\trm{\eV}}

\begin{document}

\title{Solar neutrinos: Oscillations or No-oscillations?}

\author{A.\ Yu.\ Smirnov}
\email{smirnov@mpi-hd.mpg.de}
\affiliation{Max-Planck-Institute for Nuclear Physics,
Saupfercheckweg 1, D-69117 Heidelberg, Germany}



\begin{abstract}

    The Nobel prize in physics 2015 has been awarded
``... for the discovery of neutrino oscillations which show 
that neutrinos have mass". 
While SuperKamiokande (SK), indeed, has discovered 
oscillations, SNO observed effect of
the adiabatic (almost non-oscillatory) flavor conversion
of neutrinos in the matter of the Sun. Oscillations are irrelevant 
for solar neutrinos apart from small $\nu_e$ regeneration  inside the Earth.  
Both oscillations and adiabatic conversion do not imply
masses uniquely and further studies were required to show
that non-zero neutrino masses are behind the SNO results.
Phenomena of oscillations (phase effect) and adiabatic conversion
(the MSW effect driven by the change of mixing in matter) are described in 
pedagogical way.


\end{abstract}

\maketitle

\vspace{.4cm}

\section{Introduction}

The Nobel prize in physics 2015 has been awarded to 
T. Kajita, Super-Kamiokande,  and A. B. McDonald, 
Sudbury Neutrino Observatory (SNO).
Super-Kamiokande (SK, among other things) studied properties of the atmospheric neutrinos.
The oscillatory dependence
of the number of $\mu$-like events on $L/E$ (distance over energy) has been observed which 
is the key signature of oscillations \cite{atm}. 
The SNO collaboration studied the solar neutrinos:  
the flux of the $\nu_e$ neutrinos and the total flux of 
all neutrino flavors ($\nu_e$, $\nu_\mu$, $\nu_\tau$)
have been measured via 
the charged current interactions 
and the neutral current interactions correspondingly.  
Comparing the two fluxes, SNO has established  transformation of $\nu_e$ into $\nu_\mu$
and $\nu_\tau$ \cite{sol}.

The prize has been  awarded 
``... for the discovery of neutrino oscillations, which shows
that neutrinos have mass". 
Two remarks concerning this citation are in order

\begin{itemize}

\item 
While SK has,  indeed,  discovered neutrino oscillations, 
the SNO has established,  
as we understood later, 
almost {\it non-oscillatory} adiabatic flavor conversion (the MSW effect). 
Oscillations are irrelevant 
for interpretation of the SNO results apart from small 
regeneration effect inside the Earth.

\item

Oscillations do not necessarily imply the mass.

\end{itemize}

In what follows I will explain these two points and 
give simple description of the phenomena of oscillations and adiabatic conversion. 
Comments on physics and terminology will be given in conclusion 
\footnote{This paper is based on several colloquia  delivered 
by the author during the last year.}. \\

\begin{figure}[t]
\includegraphics[width=5.0in]{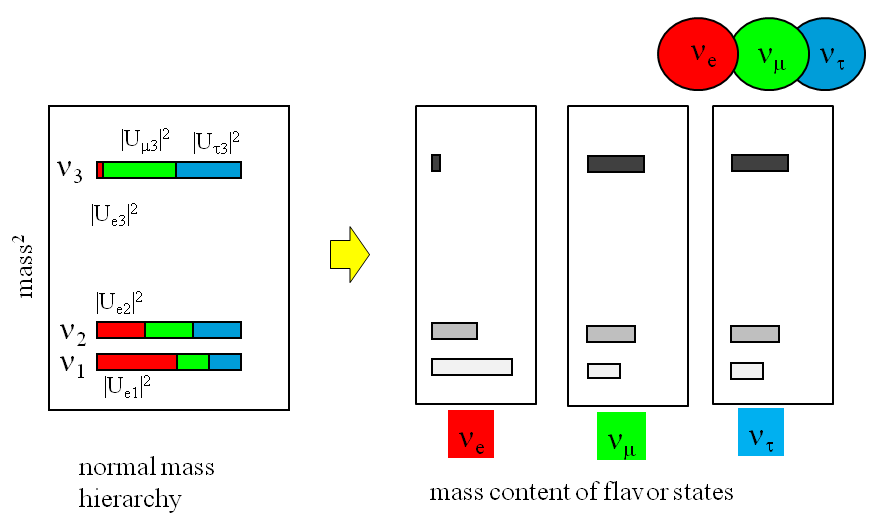}
\caption{Graphic representation of neutrino mixing. 
{\it Left panel:} neutrino mass spectrum and flavor composition of the mass eigenstates.
The mass states are shown by boxes. 
Each box contains mixture of different flavors (color parts).
Areas of colored parts  give  probabilities to find 
the corresponding flavor neutrino in a given mass state, if the area of the box is 1.  
{\it Right panel:} Mass composition of the flavor states. 
The gray-black boxes correspond to the mass states in a given flavor state.
Relative areas of the boxes give probabilities to find
the corresponding mass state in a given flavor state.}
\label{fig:gr-mix}
\end{figure}

Neutrino oscillations and adiabatic conversion 
are consequences of mixing \cite{pont}, \cite{mns}.
Graphic representation of the vacuum mixing is shown in Fig.~\ref{fig:gr-mix}. 
There are three types of neutrinos: $\nu_e$, $\nu_\mu$, $\nu_\tau$
which we refer to as  neutrinos with definite flavors.
Mixing means that the flavor neutrino states do not coincide
with the mass states  $\nu_1$, $\nu_2$,
$\nu_3$. The flavor states are combinations (mixture) of
mass states, and inversely,  the mass states are combinations
of the flavor states (see Fig. \ref{fig:gr-mix}, left). 
According to this Figure,  e.g. $\nu_2$ is composed of nearly equal amount of
$\nu_e$, $\nu_\mu$, $\nu_\tau$. In $\nu_3$
the flavor states $\nu_\mu$ and  $\nu_\tau$ are presented almost equally with
very small admixture of $\nu_e$. Therefore,  $\nu_3$ would show up
(interact) with probability $\sim 0.48$ as $\nu_\mu$,
with probability $\sim 0.5$ as $\nu_\tau$ and with probability 0.02
as $\nu_e$. If the beam of high energy $\nu_3$ 
is created,  it will produce (in the CC interactions) numbers of
$e$, $\mu$ and $\tau$ leptons with fractions $2:48:50$.

Second  aspect of mixing 
is that the flavor states
are combinations of the mass states (Fig.~\ref{fig:gr-mix}, right). 
E.g. $\nu_e$ is composed of  about 2/3 of $\nu_1$,
1/6 of $\nu_2$ and 1/6 of $\nu_3$.
A mass spectrometer studying  $\nu_\mu$ (in a ``gedadken'' experiment)  would find  
three peaks: at  values of mass $m_1$, $m_2$ and $m_3$ with intensities  
$2/3~:~1/6~:~1/6$ correspondingly.

The key point which can not be seen in this figure is that 
flavor states are  coherent combinations of the 
mass states. The mass states   $\nu_i$,  in a given flavor state
 $\nu_\alpha$ ($\alpha~ =~ e, \mu, \tau$) have definite relative phases.  

In terms of Fig. \ref{fig:gr-mix} left,    
SK has measured value of large (2-3) mass splitting and distribution of the
$\nu_\mu$ and  $\nu_\tau$ flavors (green and blue) in the third mass state $\nu_3$.
SNO has constrained the small (1-2) mass splitting and established the
distribution of the $\nu_e$ flavor (red) in  $\nu_1$ and $\nu_2$.

\section{Oscillations}

Neutrino oscillations \cite{pont} \cite{grib} are

\begin{itemize}

\item
consequence of mixing:
production and propagation of mixed states; 

\item
manifestation of interference;

\item
effect of change of the relative  phase:
increase with time and  distance of
the phase difference between the eigenstates 
of the Hamiltonian which compose a propagating mixed state. 
It is this increase of the phase that  changes the interference effect.

\end{itemize}

For simplicity 
we will consider two-neutrino mixing: $\nu_\mu$ and $\nu_\tau$ 
keeping in mind application to 
the SK results on the atmospheric neutrinos: 
\bea
\nu_\mu & = & ~~\cos \theta_{23} \nu_2 + \sin \theta_{23} \nu_3, 
\nonumber \\
\nu_\tau & = &  - \sin \theta_{23} \nu_2 + \cos \theta_{23} \nu_3 
= e^{i\pi} \sin \theta_{23} \nu_2 + \cos \theta_{23} \nu_3 . 
\label{eq:mutaumixing}
\eea
Eq. (\ref{eq:mutaumixing}) means that $\nu_\mu$, the neutrino state produced
with muon e.g., in pion decay,  
is certain coherent combination of two 
states with definite masses. The combination is characterized
by weights given by $\cos \theta_{23}$ and $\sin \theta_{23}$ 
and the relative phase.
According to equation (\ref{eq:mutaumixing})
there  is an additional ``intrinsic'' phase $\phi_{int}^{\tau} = \pi$ in the 
$\nu_\tau$ state, while and $\phi_{int}^{\mu} = 0$ in $\nu_\mu$  
\footnote{This additional phase is related to  
orthogonality of the two flavor states.
For definiteness we put it on 
$\nu_\tau$. We could put it on $\nu_\mu$ 
or, in general,  introduce the  phases 
in both states but the difference should be $\pi$.}.
Coherence means that  
the relative phases between $\nu_2$  and $\nu_3$  in $\nu_\mu$ and $\nu_\tau$ 
are not random or averaged, and they determine properties (in particular interactions) 
of these states. \\

The equations (\ref{eq:mutaumixing}) can be inverted: 
\bea
\nu_2 & = & \cos \theta_{23} \nu_\mu - \sin \theta_{23} \nu_\tau, 
\nonumber \\
\nu_3 & = &   \sin \theta_{23} \nu_\mu + \cos \theta_{23} \nu_\tau, 
\label{eq:23mixing}
\eea
which  shows  that  the mass eigenstates are combinations of the flavor states.  

\begin{figure}[h]
\includegraphics[width=2.5in]{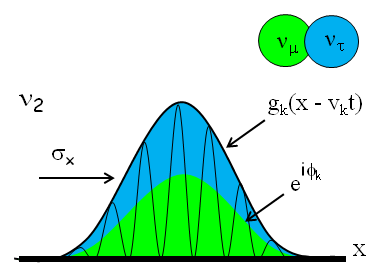}
\caption{Wave packet of the mass state $\nu_2$. The oscillatory pattern is inscribed in the 
envelope which moves with group velocity. The green and blue parts show 
the flavor composition of the mass state.}
\label{fig:wp2}
\end{figure}

Propagation of the mass states is described
by wave packets\footnote{Although for practical purpose the wave packet description 
can be avoided in most of the cases, it  is necessary to clarify the 
conceptual issues.}. 
The wave packet for the k-eigenstate ($k = 2,~3$) in the configuration space $(t, x)$ 
can be parametrized as the product of the shape and phase factors \cite{akhm}: 
\be
\psi_k = g_k (x - v_k t)~ e^{i\phi_k}. 
\ee
The shape factor  $g_k$ depends on  combination of space-time coordinates $(x - v_k t)$, 
where $v_k$ is the group velocity, and therefore describes propagation of the packet.  
$g_k$ is an envelope of the packet.  
The phase factor, $e^{i\phi_k}$,  produces an oscillatory pattern 
inscribed in the envelop (see Fig.~\ref{fig:wp2}). Here 
\be
\phi_k(x, t) = p_k x - E_k t 
\ee
is the phase with $p_k$ and $E_k$
being the mean momentum and mean energy 
in the packet (see Fig. \ref{fig:wp2}).  
The size of the WP and its shape are 
determined by the production process: by  kinematics of the elementary 
reaction  and  localization of parents (source) of neutrinos.

\begin{figure}[t]
\includegraphics[width=4.5in]{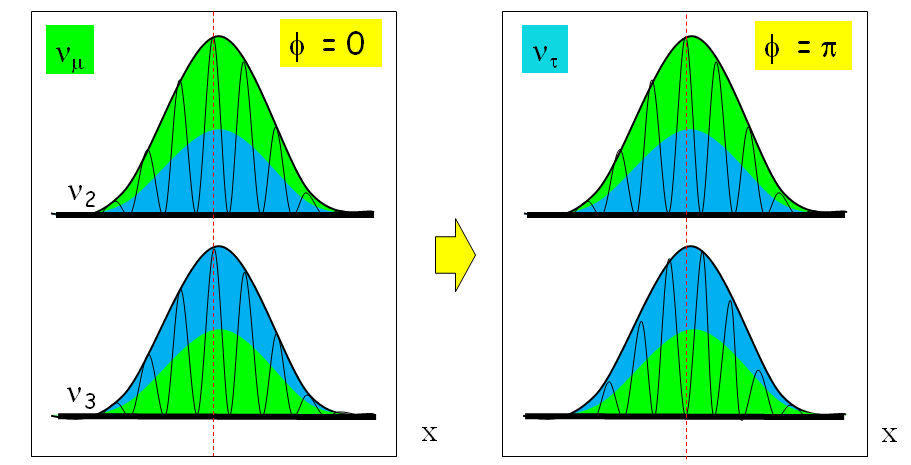}
\caption{Wave packet picture of neutrino oscillations
$\nu_\mu \rightarrow \nu_\tau$.
{\it Left panel:} picture of muon neutrino neutrino.
{\it Right panel:} tau neutrino. The two states distinguished mainly 
by the  phase difference (shift of the oscillatory patterns.) 
Each WP has muon (green) and tau (blue) parts according to Eq. (\ref{eq:23mixing}). 
}
\label{fig:wpmutau}
\end{figure}

\begin{figure}[h]
\includegraphics[width=4.5in]{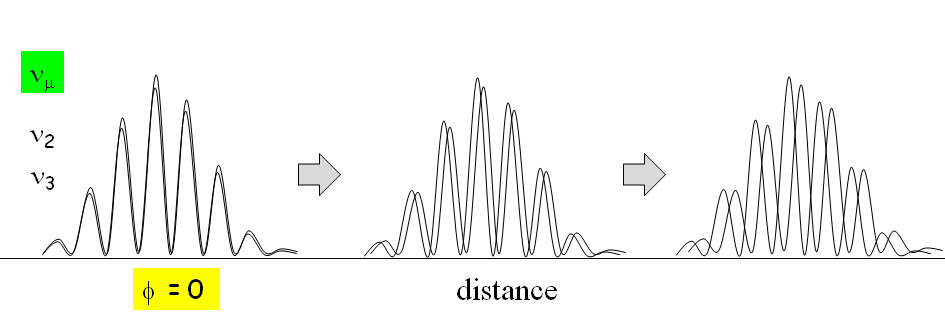}
\caption{Wave packet picture of neutrino oscillations:
shift of the phase between oscillatory factors. }
\label{fig:p-shift}
\end{figure}

In the  $2\nu-$ approximation 
propagation in vacuum of the state produced as $\nu_\mu$ (\ref{eq:mutaumixing})
is described by two wave packets which correspond to
mass states $\nu_2$ and $\nu_3$ 
(and similarly -- for $\nu_\tau$),  see Fig. \ref{fig:wpmutau}.  
In this picture there are several  simplifications 
which are not essential for physics of oscillations. 
(i)  We show one-dimensional (1D) picture; 
in 3D the WP may have spherical front. 
(ii) For visibility we show one packet under another but in reality 
they overlap in space, (iii) we show only the upper parts of the packets. 
In reality the packet is symmetric with respect to the propagation axis
(and actually,  it  occupies a complex plane). 
Parts of different WP with the same flavor (color in Fig. \ref{fig:wpmutau}) interfere 
and the result of interference depends on the phase difference. 
We assume that the WP are short enough, so that the phase difference is the same 
along the whole packet (from front to back part).

In the course of propagation 
additional phase difference between the mass eigenstates 
appears: Due to difference of masses 
the states $\nu_2$ and $\nu_3$ have different 
{\it phase velocities},  $v^{ph}_k = E_k/p_k$ ($k = 2, 3$). The latter leads to appearance 
of the phase difference (shift of the oscillatory patterns, see Fig. \ref{fig:p-shift}) during propagation: 
\be
\phi_{osc} \equiv \phi_3 - \phi_2 \approx \frac{\Delta m^2_{32} L}{2 E}   
\label{eq:osc-p}
\ee
which we call the oscillation phase.
The total phase difference between mass states in 
the $\nu_\tau$  is 
\be
\phi_\tau = \phi_{int}^{\tau} + \phi_{osc} = \pi + \phi_{osc}. 
\ee
Since $\phi_{int}^{\mu} = 0$,  the phase in the $\nu_\mu$ state equals 
\be
\phi_\mu = \phi_{osc}.
\ee

Suppose $\nu_\mu$ is produced at $t\approx 0$, $x \approx 0$ (actually we can not indicate 
exact time and space point due to
finite size of the source localization region related to the uncertainty principle.) 
In this moment the  oscillation phase $\phi_{osc}  = 0$, so that  
$\phi_\mu = 0$ and  $\phi_\tau = \pi$. 
Consequently, there is a constructive interference of the muon parts 
and destructive interference of the tau parts 
(see Fig.~\ref{fig:wpmutau}, left). 
In fact, having the same amplitudes (see Appendix ${\bf A}$),  
the tau parts cancel completely, as it should be in the $\nu_\mu$ state.

Increase of the phase difference 
with distance and time 
will change the interference picture, see Fig. \ref{fig:p-shift}.
For $\phi_{osc}  \neq 0$ the tau parts will not cancel completely which means that 
$\nu_\tau$ component appears in the neutrino state originally produced as $\nu_\mu$. 
When $\phi_{osc} = \pi$ (corresponding to the strongest 
deviation from the initial state) we have $\phi_\mu = \pi$ and  $\phi_\tau = 2\pi$. 
This leads to the destructive interference of the muon parts of two packets 
and constructive interference of the tau parts (Fig. \ref{fig:wpmutau}, right). 
Thus,  the originally produced muon neutrino
is  converted partially or completely (if mixing is maximal) to the tau neutrino. 
The neutrino system  returns back to the initial  flavor state when
$\phi_{osc} = 2\pi$. According to (\ref{eq:osc-p})
the corresponding distance -- the oscillation length --  
equals 
\be
l_\nu = L(2\pi) = \frac{4 \pi E}{\Delta m^2}. 
\ee

From Fig.~\ref{fig:wpmutau} one can immediately obtain 
the $\nu_\mu \rightarrow \nu_\mu$  
oscillation probability (see Appendix ${\bf A}$):  
\be
P_{\mu \mu} = 1- P_{\mu \tau} =
1 -\sin^2 2\theta_{23} \sin^2 \frac{\Delta m_{32}^2 L}{4E},
\label{eq:surv}
\ee
where we used explicit expression for $\phi_{osc}$ (\ref{eq:osc-p}). 
It is this phenomenon that happens in atmospheric 
neutrinos and was detected by Super-Kamiokande.






\section{No-oscillations}

\subsection{SNO results and their interpretation} 

In the case of solar neutrinos we deal with mixing of $\nu_e$ and $\nu_a$   
(the latter is certain combination of the muon and tau neutrinos): 
\bea
\nu_e & = & \cos \theta_{12} \nu_1 + \sin \theta_{12} \nu_2, 
\nonumber \\
\nu_a & = &  - \sin \theta_{12} \nu_1 + \cos \theta_{12} \nu_2.  
\label{eq:eamixing}
\eea
Here $\theta_{12}$ is the 1-2 vacuum mixing angle,  $\nu_1$ and  $\nu_2$
are the eigenstates with mass squared splitting $\Delta m_{21}^2$.

SNO has measured the flux of $\nu_e$, $\Phi_e$, detecting  
the charged current interactions and 
the total flux of all active neutrinos, $\Phi_{NC}$, 
detecting the neutral current  events. Since only $\nu_e$ are produced in the Sun, 
the ratio of the fluxes \cite{sol}:  
\begin{equation}
    P_{SNO} \equiv  \frac{\Phi_{e}}{\Phi_{NC}} = 0.340 \pm 0.023 ~^{+0.029}_{-0.031},  
\label{eq:sno}
\end{equation}
being smaller than 1 implies an  appearance of the 
$\nu_\mu$, $\nu_\tau$ fluxes in originally produced  $\nu_e$ flux, 
that is $\nu_e \rightarrow \nu_\mu, ~\nu_\tau$ transformation. 
The ratio 
$P_{SNO}$ gives the survival probability of the electron neutrinos. 
It turns out that value in (\ref{eq:sno}) is close to the value of the 
$\nu_e-$ survival probability of the so called {\it non-oscillatory transition} \cite{adiab}: 
\begin{equation}
    P_{non-osc} = \sin^2 \theta_{12} = 0.31, 
\label{eq:non}
\end{equation}
(see later). 
The difference of values in  (\ref{eq:sno}) and (\ref{eq:non}), 
$$
P_{SNO} - P_{non-osc} = 0.03, 
$$
is due to small averaged oscillation effect in the Sun and  the $\nu_e$ {\it regeneration} 
in the matter of the  Earth (see review \cite{maltoni}). 

Interpretation of the SNO result  is {\it the adiabatic flavor 
conversion} of $\nu_e$  in matter with varying density 
(the MSW effect) \cite{wolf,wolf2,ms2,adiab}. 
Complete expression for the survival probability, which reproduces (\ref{eq:sno}), is  
\be 
P_{SNO} = \sin^2 \theta_{12} + \cos 2 \theta_{12} \cos^2 \theta_{12}^{m 0} +
f_{reg}.    
\label{eq:exact-f}
\ee
Here the first term is the contribution from 
the non-oscillatory transition. 
The second one is the effect of averaged oscillations in the Sun:
\be
P_{osc} = \cos 2 \theta_{12} \cos^2 \theta_{12}^{m 0} \approx   0.015,  
\ee
where $\theta_m^0$ is the  mixing angle in matter
in the production point. Detailed study of this contribution and its dependence 
on energy has been performed in \cite{pedro}, \cite{park}. 
Finally, 
\be
f_{reg} \approx  0.015 
\ee
is the so called  $\nu_e-$regeneration factor in the Earth. 
The regeneration  is due to (averaged) oscillations of the mass states in matter  
which do not exist in vacuum. 

Thus, the effect observed by SNO is mainly the non-oscillatory transition 
with less than $10 \%$ contribution from the averaged oscillations. 
Actually, there is no much sense to speak 
about oscillations: for the SNO energies 
the coherence between the eigenstates is lost 
at distances comparable with solar radius (see below). \\

It is not possible to obtain $P_{non-osc}$ from the oscillation 
formula.  Recall that the averaged survival $2\nu$ probability in vacuum is 
$P = 1 -  0.5 \sin^2 2\theta_{12} \approx 0.58$, and in view of smallness 
of the 1-3 mixing it can never be smaller than $\approx 0.5$.   
The reason is that physics (dynamics) is different.
To explain this result let us recall some basics of 
the matter effects.

\subsection{Refraction}

At low energies inelastic interactions 
of neutrinos can be neglected. 
For these neutrinos the Earth, the Sun, other stars  
look  like transparent balls of glass for the light. 
Elastic forward  scattering produces the refraction \cite{wolf}  
--  dominant phenomena 
described by the refraction index $n$. 
The deviation of $n$ from unity
is very small. E.g.  for 10 MeV neutrino energy:
$|n - 1| = 10^{-20}$ in the Earth and  $|n - 1| = 10^{-18}$
in the center of the Sun. 
So,  what happens at low energies is extremely 
small reflection from
sharp borders between layers with different densities and bending of trajectories 
in non-uniform medium with smooth density change. 

The matter effect can be equivalently described by 
the potential $V$:  
\be
n - 1 = \frac{V}{p}. 
\ee
Not potential itself but difference of potentials has physical
meaning. For single neutrino species reflection and bending of trajectory  
are related to change of $V$ in space. 
For system of two and more mixed neutrinos the difference of potentials 
can be realized in different way -- 
due to the  difference of interactions of two 
neutrinos \cite{wolf}.   
For $\nu_e$ and $\nu_a$ the difference of potentials in
usual medium equals 
$V = V_e - V_\mu  = \sqrt{2}G_F n_e$. 
Here $G_F$ is the Fermi coupling constant and  $n_e$ is the number density of 
electrons. 
The inverse of the potential,  $l_0 \equiv 2\pi/V$,  gives the 
refraction length which determines spatial scales of the phenomena.

Amazing fact 
is that the energies of neutrinos from natural sources (MeV - GeV range),    
the radius of the Earth $R_E$ and $\Delta m_{21}^2$ 
(which comes from some new physics at very high mass scales) 
turn out to be  related in such a way that 
\be
\frac{\Delta m^2_{21}}{2 E} \sim V \sim 1/R_E. 
\ee
For neutrinos from  pion and muon decays in flight 
($\sim$ few hundreds MeV  - GeV) one should use  the 1-3 mass splitting  $\Delta m^2_{31}$   
instead of $\Delta m^2_{21}$.
It is this ``conspiracy''  of small quantities that is behind the fact that we can
observe oscillations and various matter effects in spite of smallness of $V$.   

\subsection{Mixing in matter}

\begin{figure}[t]
\includegraphics[width=4.0in]{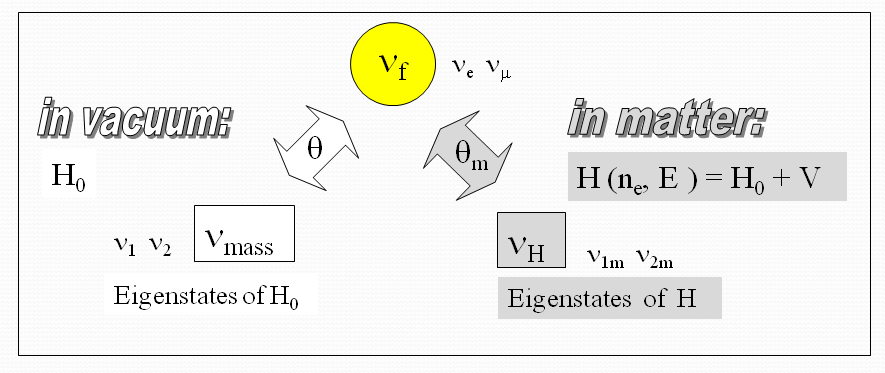}
\caption{Mixing in vacuum and in matter. 
}
\label{fig:mixv-m}
\end{figure}

In general,  the flavor mixing is defined with respect to the eigenstates 
of the Hamiltonian (Fig. \ref{fig:mixv-m}). 
In vacuum we have the Hamiltonian $H_0$ whose eigenstates   
are  the mass states $\nu_{mass}$. 
Mixing in vacuum connects the flavor states $\nu_f$ and $\nu_{mass}$  
as in eq. (\ref{eq:eamixing}).  
In matter, the total Hamiltonian differs from $H_0$. We should add the potential $V$ 
(the matrix of potentials), 
so that $H = H_0 + V$.  Therefore 
the eigenstates of $H$, $\nu_m$, differ from the eigenstates of $H_0$:  
$\nu_m \neq \nu$.   
So, instead of (\ref{eq:eamixing}) we have 
\bea
\nu_e & = & \cos \theta_{12}^m ~\nu_{1m} + \sin \theta_{12}^m \nu_{2m},
\nonumber \\
\nu_a & = &  - \sin \theta_{12}^m \nu_{1m} + \cos \theta_{12}^m \nu_{2m},  
\label{eq:eamixing-m}
\eea
and mixing angle in matter $\theta_m$
differs from mixing in vacuum: $\theta_m \neq \theta$. 
Let us reiterate: while the flavor states are the same  both in vacuum and in  
matter,  the eigenstates are different, and correspondingly,  
the mixing angles are different.  

Eq. (\ref{eq:eamixing-m}) can be inverted, expressing the eigenstates 
in terms of the flavor states: 
\bea
\nu_{1m} & = & \cos \theta_{12}^m ~\nu_{e} - \sin \theta_{12}^m \nu_{a},
\nonumber \\
\nu_{2m} & = &   \sin \theta_{12}^m \nu_{e} + \cos \theta_{12}^m \nu_{a}.
\label{eq:eamixing-m1}
\eea
According to these equations mixing (by definition) determines flavor 
composition of the eigenstates. 
Mixing angle fixes (uniquely) the flavor of the eigenstates of the Hamiltonian.   
Mixing is equivalent to flavor composition. In matter mixing depends on density. 
When density changes the mixing changes and therefore flavor of 
eigenstates changes. 

The vacuum mixing angle  is the fundamental parameter of the Hamiltonian. 
In matter the 
mixing angle  becomes a {\it dynamical variable}.
Since $H_0  = H_0(E)$  and $V = V(n)$ we have 
\be
\theta_m = \theta_m (n(t), E).  
\ee
The mixing angle is not a constant anymore and 
dependence of the  angle on the density and energy  leads to various new 
phenomena. (Introduction of notions of mixing and eigenstates in matter is very  
useful both for solution of the evolution equation 
and for understanding physics; further comments are in Appendix ${\bf B}$.)
Notice that also $E$ may depend on time as is realized, e.g.,   
due to the redshift in the Universe.\\

Dependence of $\sin^2 2 \theta_{m}$ on energy or density has resonance character. 
In the resonance $\sin^2 2 \theta_{m} = 1$ or 
\be
\theta_m (E, n) = \frac{\pi}{4}, 
\ee
and this is satisfied under the resonance condition:  
\be
V(n) = \frac{\Delta m_{21}^2}{2 E} \cos 2 \theta_{12}. 
\label{eq:res-cond}
\ee
This condition determines
the resonance density for  a given energy $E$: $n_R = n_R(E)$, 
or resonance energy in medium with 
a given density $n$: $E_R = E_R(n)$. 
The resonance energy and resonance density determine scales of various phenomena. 

In matter with constant density 
$\theta_{m}$ plays the same role as $\theta$ in vacuum 
\footnote{By the way,  $\theta$ can also be considered as the angle in medium (average of the classical field), 
keeping in mind that masses and mixing of neutrinos are generated by their interactions  
with the vacuum expectation value 
of the Higgs field(s).}. In particular,  
$\sin^2 2 \theta_{m}$ gives the 
depth of oscillations. 
At the resonance energy the depth becomes maximal:  
$\sin^2 2 \theta_{m} = 1$. This   
phenomenon was called {\it the resonance enhancement} of oscillations \cite{ms2}.

In the case of the standard neutrino interactions   
the specific dependence of mixing on $V$  is determined by the fact that 
the potential  does not change flavor being ``flavor-diagonal".  
Scattering which produces  this potential does not change flavor. 
So, in the Hamiltonian it appears in diagonal elements,  
while the flavor is changed by the off-diagonal 
term $\sin 2 \theta  \Delta m^2/4E$ which does not depend on density.  
Therefore at very high densities --  
much above the resonance  $V \gg V(n_R)$, 
the diagonal elements  dominate and 
therefore mixing is suppressed. This has two realizations 
depending on signs of $\Delta m^2$ and  $V$ (and $V$ 
has opposite signs for neutrino  and antineutrinos):  
$\theta_{12}^{m} \rightarrow \pi/2$ ($n \gg n_R$) in the resonance channel,  
where the condition (\ref{eq:res-cond})  is satisfied,  and   
$\theta_{12}^{m} \rightarrow 0$  which occurs in the non-resonance channel.  
In the resonance channel with decrease of density the angle $\theta_{12}^{m}$ decreases 
from $\pi/2$: 
it becomes $\pi/4$ in the resonance and then approaches 
the vacuum value $\theta_{m}  \approx  \theta$ when $n \ll n_R$.

\subsection{Adiabatic conversion}

Adiabatic conversion is realized in matter with 
slowly varying density \cite{wolf2, adiab}. 
As we discussed in the previous section  mixing connects uniquely 
the flavor state with the eigenstates. 
So, dynamics of flavor transformations is reduced to the dynamics of 
the eigenstates. 
Transitions between the eigenstates $\nu_{1m} \leftrightarrow \nu_{2m}$  
are governed by the adiabaticity condition
which involves density gradient on the way of neutrinos. 
If density changes slowly enough this  condition is fulfilled (see Appendix ${\bf C}$), 
and consequently,  transitions between the eigenstates are negligible. 
The absence of the 
\be
\nu_{2m}  \leftrightarrow \nu_{1m}
\ee
transitions is the essence of the  adiabatic propagation.\\ 

\begin{figure}[h!]
\includegraphics[width=5.0in]{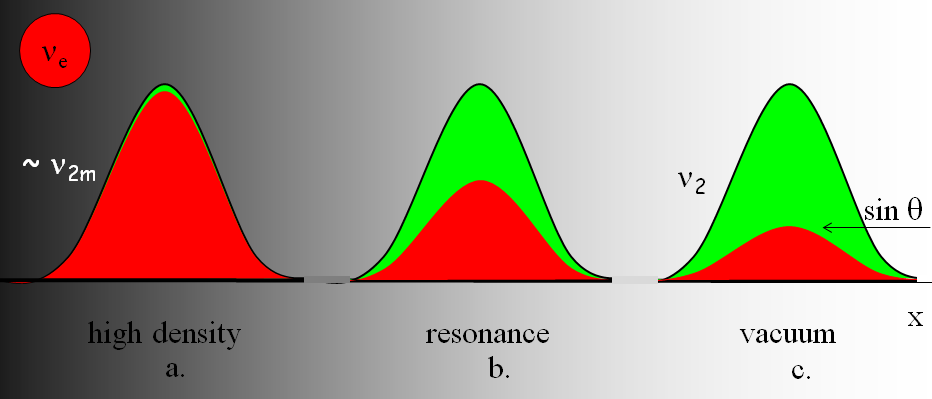}
\caption{Wave packet picture of the non-oscillatory adiabatic conversion of
the electron neutrino. Initial density $n \gg n_R$. Shown are 
snapshots of the propagating neutrino state for three different densities. 
The neutrino state $\nu_e$ consists essentially of a single 
eigenstate $\nu_{2m}$. The flavor of this eigenstate determined 
by mixing angle changes according 
to the density change. The irrelevant oscillatory pattern 
(as in Figs. \ref{fig:wp2}, \ref{fig:wpmutau}) is not shown here.  
}
\label{fig:non-osc}
\end{figure}

Let us describe now the non-oscillatory transitions. 
The benchmark point here is again the resonance density $n_R$. 
Suppose the electron  neutrinos are produced at very high densities  
$n \gg n_R$. As discussed in sect. IIIC, in this case mixing is very strongly suppressed and  
in the resonance channel  $\theta_{12}^{m 0} \approx \pi/2$. 
This means, according to eq. (\ref{eq:eamixing-m}),   
that the electron neutrino is essentially composed of single 
eigenstate $\nu_e \approx\nu_{2m}$. In turn, as follows from (\ref{eq:eamixing-m1}) 
this eigenstate has mainly the  electron flavor:  $\nu_{2m}  \approx \nu_e$.  
The admixture of the second eigenstate is strongly suppressed 
and can be neglected (see  Fig. \ref{fig:non-osc}a). 
If single eigenstate  $\nu_{2m}$ propagates,  
there is no interference since simply there nothing 
to interfere with. Consequently, there is no oscillations 
in principle. The phase of this state is irrelevant 
(only the difference of phases has physical meaning).

If the density changes on the way of neutrino,  then mixing $\theta_{12}^{m}$ 
changes, and consequently, flavor of the eigenstate  $\nu_{2m}$ 
changes. With decrease of density  the $\nu_a$ component in 
$\nu_{2m}$ increases.  In resonance $\nu_{2m}$ has equal fraction of $\nu_e$ and 
$\nu_a$ (see Fig. \ref{fig:non-osc}d), 
and then fraction of $\nu_a$ becomes bigger than the one of  $\nu_e$ approaching 
the vacuum value. 
At low density, the flavor of the eigenstate is  determined 
by the vacuum mixing (Fig. \ref{fig:non-osc}c). 

When density changes slowly (adiabatically) another eigenstate 
$\nu_{1m}$ is not  produced. So, during whole the evolution 
$\nu \approx \nu_{2m}$, and consequently,  change of
flavor of the whole state follows the flavor of 
$\nu_{2m}$ and the latter, in turn,  follows the density variations. This is realization 
of the adiabatic non-oscillatory regime:  
\be
\nu_e \approx \nu_{2m}  \rightarrow \nu_{2}. 
\ee
Now we can obtain  the probability 
of the transition (survival probability) immediately: 
\be
P_{ee} \approx  |\langle\nu_e | \nu_2 \rangle|^2 =
\sin^2 \theta_{12}. 
\ee

\begin{figure}[h!]
\includegraphics[width=5.0in]{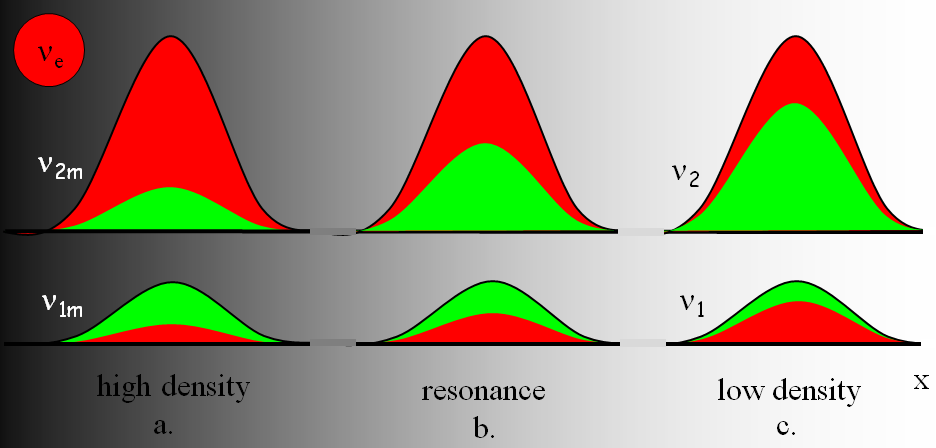}
\caption{Wave packet picture of the adiabatic
conversion of the electron neutrino in general case.  
Sizes of the WP do not change; flavors of the eigenstates change according 
to density change. Interference of the same flavor parts 
takes place. The oscillatory pattern is not shown: the wave packets of 
$\nu_1$ and $\nu_2$ eventually separate and the interference terminates.}
\label{fig:adiab-osc}
\end{figure}

If initial density is not very large,  then 
the produced $\nu_e$ is  composed of two eigenstates -- 
the admixture of $\nu_{1m}$ can not be neglected,    
see Fig.~\ref{fig:adiab-osc}a. 
The admixtures are determined by mixing in the initial moment (production point).  
In this case  apart from change  of flavors of individual 
eigenstates  $\nu_{1m}$ and $\nu_{2m}$ also interference  takes place and so 
an interplay of the adiabatic conversion and oscillations should be observed.

\begin{figure}[h]
\includegraphics[width=4.5in]{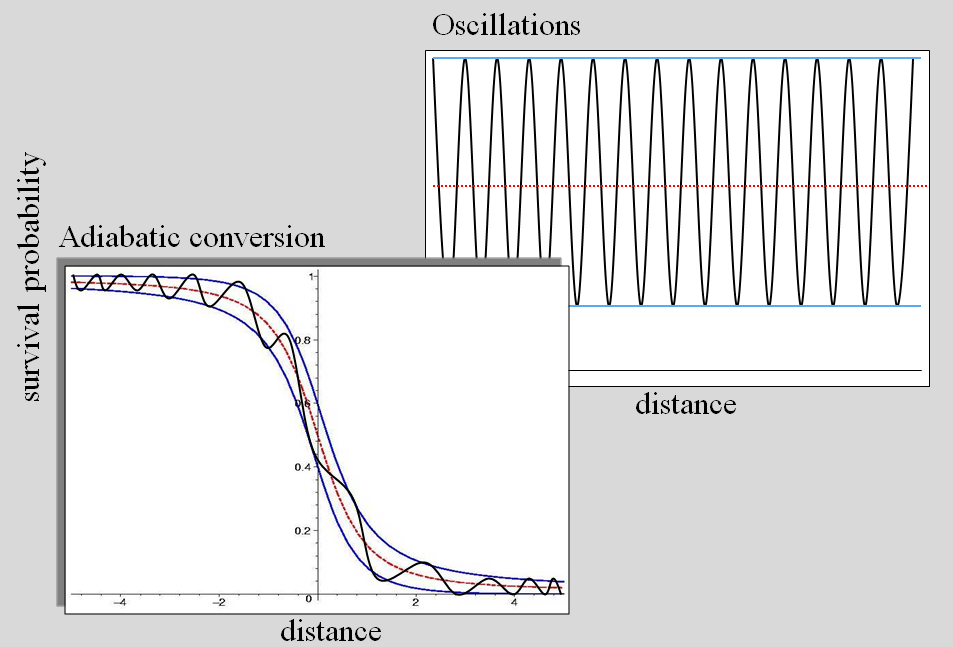}
\caption{Spatial picture of oscillations (upper panel) and adiabatic conversion (lower panel). 
Shown are the dependence of the survival probabilities on distance.}
\label{fig:spatial-ad}
\end{figure}

Again,  if the density changes slowly the transitions 
$\nu_{2m}  \leftrightarrow \nu_{1m}$ 
are suppressed,  and therefore 
the amplitudes (shape factors) of the wave packets 
of $\nu_{2m}$ and $\nu_{1m}$ do not change. 
Flavors of the eigenstates being determined by 
mixing angle $\theta_{12}^m$ follow the density change. 
The dependence of the survival probability on distance  
for the adiabatic conversion is shown in 
the lower panel of Fig. \ref{fig:spatial-ad}, 
being compared with spatial picture of oscillations (the upper panel).\\ 

\begin{figure}[h]
\includegraphics[width=4.5in]{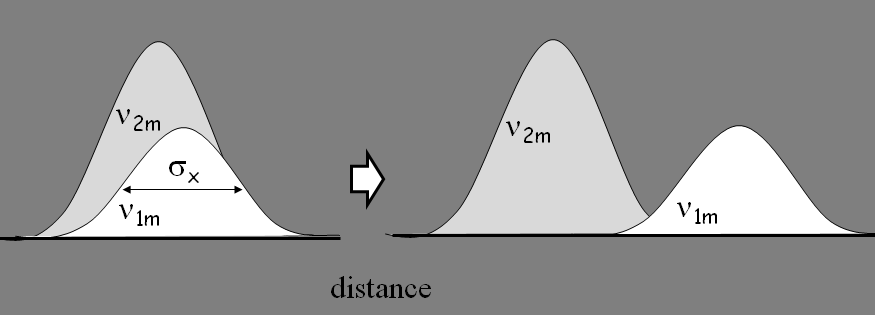}
\caption{Loss of propagation coherence due to spatial separation of the wave packets.}
\label{fig:separation}
\end{figure}

Even in the presence of both eigenstates 
for solar neutrinos the oscillations are irrelevant.  
Indeed, in the configuration space   
loss of the propagation coherence occurs: 
due to difference of the group velocities (they should be computed in matter) 
 the wave packets  which correspond to two eigenstates 
shift with respect to  
each other and eventually separate (Fig. \ref{fig:separation}).  
The difference of  group velocities in vacuum is  
\be
\Delta v_{gr}  =  \frac{\Delta m^2}{2E^2}. 
\ee
The shift equals $\Delta v_{gr} L$. 
Therefore complete separation of the packets occurs at the distance $L_{coh}$ 
(coherence length) determined by 
\be
\Delta v_{gr} L_{coh}  =  \frac{\Delta m^2L}{2E^2 } = \sigma_x,  
\ee
where $\sigma_x$ is the size of the wave packet. For solar neutrinos depending on 
energy $L_{coh}$ varies from few hundreds of kilometers to several solar radius's
\footnote{This loss of coherence can not be restored at the 
detection,  since it would require too long coherence time of 
detection, or equivalently,  unachievable  energy resolution.}.  \\

Conversion of the solar neutrinos can be viewed as the incoherent production 
and propagation of the eigenstates in matter, and  change of flavors of these eigenstates 
in course of propagation according to density change. 
Admixtures (weights) of the eigenstates in a given propagating state do not change being 
determined by $\theta_{12}^{m0}$ -- the mixing angle in matter in the production point. 
The accuracy of this description is $\sim 10^{-4}$. 

The survival probability can be obtained from the Fig. \ref{fig:adiab-osc} immediately.
The amplitudes of the packets in initial moment,  and consequently, 
in any other moment (due to adiabaticity) are determined
by mixing in the initial state:  $\cos \theta_{12}^{m0}$ for $\nu_{1m}$ and
$\sin \theta_{12}^{m0}$ for $\nu_{2m}$.
Due to adiabaticity and loss of coherence  the two wave packets evolve 
independently and effects of these two packets should sum up  in the
probability. The contribution to the probability to find $\nu_e$ from  $\nu_{1m}$ which transforms to $\nu_{1}$ is
$(\cos \theta_{12} \cos \theta_{12}^{m0})^2$ where $\cos \theta_{12}$ gives fraction of $\nu_e$ in $\nu_{1}$.  
Similarly, the contribution from  $\nu_{2m}$ is 
$(\sin \theta_{12} \sin \theta_{12}^{m0})^2$. The sum of the contributions reproduces the
first and second  terms in Eq. (\ref{eq:exact-f}). \\


\begin{figure}[h]
\includegraphics[width=4.5in]{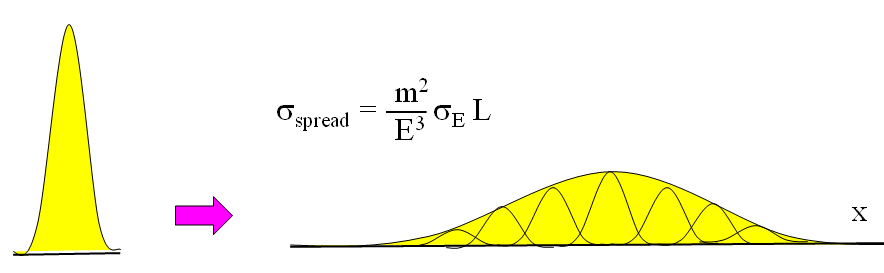}
\caption{Spread of the individual wave packet of the eigenstate with mass $m$. 
Different parts of the expanded WP of the size $\sigma_x$ become effectively incoherent.}
\label{fig:spread}
\end{figure}

Let us mention that there  is the  third effect of propagation: 
spread  
of individual  wave packets in space. 
The spread  is related to the presence of different momenta  
in a given wave packet: so that parts of the WP  with 
higher energy will propagate faster (see Fig. \ref{fig:spread},  and details in \cite{kersten}).
It can be shown that spread of the WP does not change the coherence condition. 

Bottom line: SNO had observed effect of the adiabatic conversion 
and loss of coherence.

\subsection{Oscillations versus adiabatic conversion}

Let us summarize the  differences between  the oscillations and adiabatic conversion. 
Neutrino oscillations are manifestation of interference which changes in  space/time. 
The interference is  determined by the 
phase difference between the two (or more) eigenstates of the Hamiltonian.

In the case of the non-oscillatory transition 
a produced neutrino state consists of a single eigenstate.  
Consequently,  there is no interference, no phase difference, 
and no oscillations.  

In pure form oscillations occur in vacuum and matter with constant density. 
In contrast, the adiabatic flavor conversion is the effect of propagation of neutrino in medium with
slowly changing density.

Two different degrees of freedom are involved 
in oscillations and conversion:

\begin{itemize}

\item
the phase $\phi_{osc}$  -- in the case of oscillations.   It changes the interference picture.
The phase $\phi_{osc}$ is the key dynamical  variable;  mixing does not change.

\item 
mixing angle $\theta_m $ -- in the case of 
adiabatic conversion.  Here the  flavor of a neutrino state determined by $\theta_m(n)$ follows
the density change on the way of neutrinos.
The phase is irrelevant. The process is not periodic and irreversible for solar neutrinos.

\end{itemize}

Non-oscillatory transition is extreme case of the adiabatic
conversion when initial density is much larger
than the resonance density. This is realized approximately
for high energy part of the solar neutrino spectrum.

Periodic (or quasi-periodic) $L/E$ dependence of the probability is 
the main signature of oscillations.  
Oscillations show up as oscillatory dependence of the flavor
of neutrino state as function of distance $L$ or
$L/E$,  in general.
Adiabatic conversion is independent of  spatial scales 
- the corresponding probabilities do not depend on distance.

\begin{figure}[h]
\includegraphics[width=4.0in]{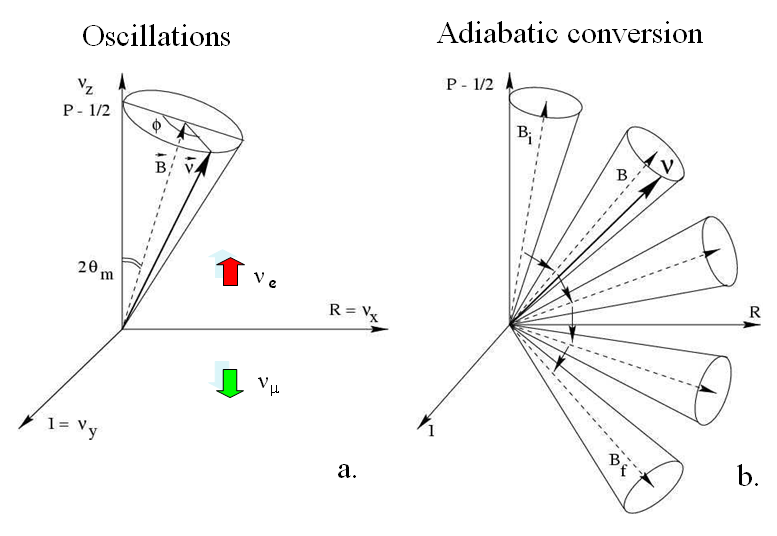}
\includegraphics[width=2.3in]{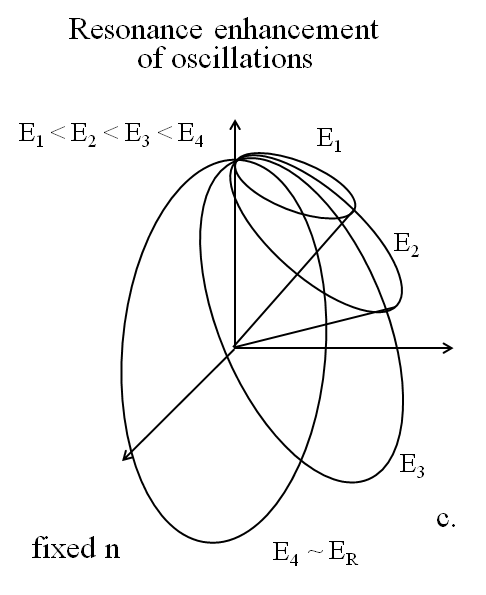}
\caption{Graphic representation of the neutrino oscillations (a),   
adiabatic conversion (b), and the resonance enhancement of oscillations (c). 
The non-oscillatory transition corresponds to b) with very small cone opening angle, so that the cone essentially 
coincides with the axis.  
In the case c) for different energies neutrino vector
moves on the surface of the cone with different directions of the axis and
cone angle. Shown are cones of rotation for different energies. 
}
\label{fig:os-ad}
\end{figure}

In Fig. \ref{fig:os-ad} we show  graphic representations 
of the neutrino oscillations and
adiabatic conversion which are based on analogy with
the electron spin precession in the magnetic field. 
Neutrino polarization vector in flavor space (``spin'') 
is moving in the flavor space around the ``eigenstate axis" 
(magnetic field) whose direction is determined by the mixing angle $2\theta_m$.  
Oscillations are equivalent to the precession of the neutrino polarization vector
around fixed axis, Fig. \ref{fig:os-ad}a. 
Oscillation probability
is determined by projection of the neutrino vector on the axis $z$.
The  direction up  of the neutrino vector corresponds to
the $\nu_e$, direction down -- to $\nu_a$.
Adiabatic conversion is driven by rotation of the cone itself, 
{\it i.e.} change of direction of the magnetic field (cone axis)
according to change of the mixing angle,  Fig. \ref{fig:os-ad}b. 
Due to adiabaticity the cone opening  angle does not change
and therefore the neutrino vector follow rotation of axis.


\section{``... which shows that neutrinos have mass"}

\subsection{SNO, oscillations and KamLAND}

SNO has established the transformations of $\nu_e$ to $\nu_\mu$, $\nu_\tau$. 
Within the experimental  error-bars the effects does not depend on energy.  
No $L/E$ dependence 
has been observed and mechanism of transformation 
has not been identified. After SNO publications a number of 
solutions of the solar neutrino problem still  existed: 
LMA MSW and   LOW MSW conversion,  
resonant spin-flavor precession, 
Lorentz symmetry violation, decoherence,  neutrino decay, {\it etc}.
Very good description of the data has been provided 
by the so called non-standard neutrino 
interactions with zero neutrino masses. 

It is the  KamLAND (Kamioka Large Antineutrino detector) \cite{kamland}
that has selected unique solution of the solar neutrino problem 
(in assumption of the CPT invariance) and showed that non-zero neutrino mass 
is behind the SNO result.  
KamLAND  studied the antineutrino fluxes from  many atomic reactors 
with average baseline about 180 km. 
The $L/E$ dependence of the survival probability has been observed. 
Extracted values of $\Delta m^2_{21}$ and $\theta_{12}$ were in agreement 
with LMA MSW solution.   


Solar neutrino experiments and KamLAND have completely different environments: 
Solar neutrinos propagate in matter with varying density, then in vacuum 
and finally in matter of the Earth. In KamLAND we deal essentially 
with oscillations in vacuum (matter effect is present but very small).
Coincidence of the parameters $\Delta m^2_{21}$,  $\theta_{12}$ 
determined in solar experiments and KamLAND  
had a number of implications: 

- confirmation of CPT  invariance; 

- correctness of theory of neutrino oscillations in vacuum and in matter,  and 
of adiabatic conversion; 

- selection of  unique solution  of the solar neutrino problem,

- strong indication  that neutrino mass is behind oscillations 
and adiabatic conversion.

15 years after  with more data collected we see some difference  
of $\Delta m^2_{21}$ extracted from the solar  data and KamLAND.  
The significance of this difference is slightly bigger than 
$2\sigma$. It can be just statistical fluctuation or some systematics, but may turn out 
to be effect of new physics.  



\subsection{Oscillations and mass}

Oscillations do not need the mass. 
Recall that  it was the subject of the classical 
Wolfenstein's paper \cite{wolf} to show that oscillations can proceed 
for massless neutrinos. This requires, however,  introduction of 
the non-standard interactions of neutrinos which lead to 
non-diagonal potentials in the flavor basis and therefore produce mixing. 

In oscillations 
we test the dispersion relations,  
that is, the relations between the energy and momentum,   
and not masses immediately. Oscillations are induced because of 
difference of  dispersion of neutrino components that compose a mixed 
state. In vacuum the relation reads
\be
E = \sqrt{p^2 + m^2} \approx p + \frac{m^2}{2 p}. 
\label{eq:disp}
\ee
The mass squared enters  here  (so,  the  chirality flips twice), 
and  eventually there  is no change of chirality:  
$\nu_L \rightarrow \nu_R \rightarrow \nu_L$. Therefore  
V, A interaction 
with medium can reproduce effect of mass.   

In the presence of matter, the dispersion relation 
becomes 
\be
E = \sqrt{p^2 + m^2} + V \approx p + \frac{m^2}{2 p}
+ V. 
\label{eq:disp2}
\ee
The matter potential can be considered as 
an effective mass. Indeed, for massless neutrinos Eq. (\ref{eq:disp2}) can be rewritten as 
$E = \sqrt{p^2 + m_{eff}^2}$,   where 
\be
m^2_{eff} =  2pV. 
\label{eq:disp3}
\ee
The effective mass has momentum dependence which allows to 
disentangle it from true mass. \\

It is consistency of results of many experiments 
in wide energy ranges and different environments: 
vacuum, matter with different density profiles 
that makes explanation of data 
without mass almost impossible.

In this connection one may wonder which type 
of experiment/measurement can uniquely  identify 
the true mass? Let us mention three possibilities: 

\begin{itemize}

\item
Kinematical measurements: distortion of the beta 
decay  spectrum near the end point.  
Notice that  similar effect can be produced 
if  a degenerate sea of neutrinos exists 
which blocks neutrino emission near the end point.  

\item 
Detection of neutrinoless double beta decay which is the test of the Majorana 
neutrino mass. Here complications are related to possible contributions 
to the  decay from new $L$-violating interactions. 

\item 
Cosmology is sensitive to the sum 
of neutrino masses, and in future it will be sensitive to even individual masses. 
Here the problem is with degeneracy of neutrino mass and  cosmological parameters. 

\end{itemize}

\section{In conclusion}

In some cases (for historical or other reasons) 
terminology does not correspond  to real physics.
In most of the cases we understand difference and what is behind.
Still  bad terminology can be misleading 
producing wrong  physics interpretations.

Calling the two different effects (oscillations and adiabatic conversion) 
just oscillations is simpler and shorter. 
In fact, both the oscillations and adiabatic conversion can be consequences of
neutrino mass and mixing. Also neutrino decay is a consequence
of mass and mixing, but  we do not call it oscillations.
Partly it was our fault with Mikheyev:  
In our early  publications 
we described two different matter effects under the same name: 

\begin{itemize}

\item 

Resonance enhancement of oscillations which takes place
in matter with constant (quasi constant) density, like mantle of the Earth. 
Here the phase is crucial. 
Graphic representation of this effect is given in 
Fig. \ref{fig:os-ad}c.  The effect hopefully will be
observed by PINGU, ORCA experiments, and will allow us to determine 
the neutrino mass hierarchy.  

\item 
Adiabatic conversion of neutrinos 
which (as we have discussed) takes place 
in matter with slowly changing density (the Sun, supernovae).
Resonance is important also here determining strength of transitions: 
strong transitions occur in the resonance channel when e.g. neutrinos are produced 
at densities much above the resonance one, cross the resonance layer and then exit matter 
at densities much below the resonance density.

\end{itemize}

In January 1986 at the Moriond workshop A. Messiah
(he gave the talk \cite{messiah}) asked me:
`` why do you call effect that happens in the Sun the resonance
oscillations? It has nothing to do with oscillations, 
I will call it the  MSW effect".
My reply was ``yes, I agree, we simply did know how to call it. 
I will explain and correct this in my future  talks and publications''.
Messiah's answer was surprising:
``No way..., now this confusion will stay forever". That time I could not believe him.
I have published series of papers,
delivered review talks, lectures in which I was trying to explain,
fix terminology, {\it etc.}. 
All this has been described in details in the 
talk at Nobel symposium \cite{s-nob},  and  
for recent review see \cite{maltoni}.

Ideally terminology should reflect and follow our understanding
of the subject. Deeper understanding may require a change
or modification of terminology. At the same time changing terminology is very delicate 
thing and can be done with great care. \\


In conclusion, the answer to the question in the title of the paper is \\

\begin{center}

``Solar neutrinos: Almost No-oscillations''. \\

\end{center}

\begin{figure}[h]
\includegraphics[width=5in]{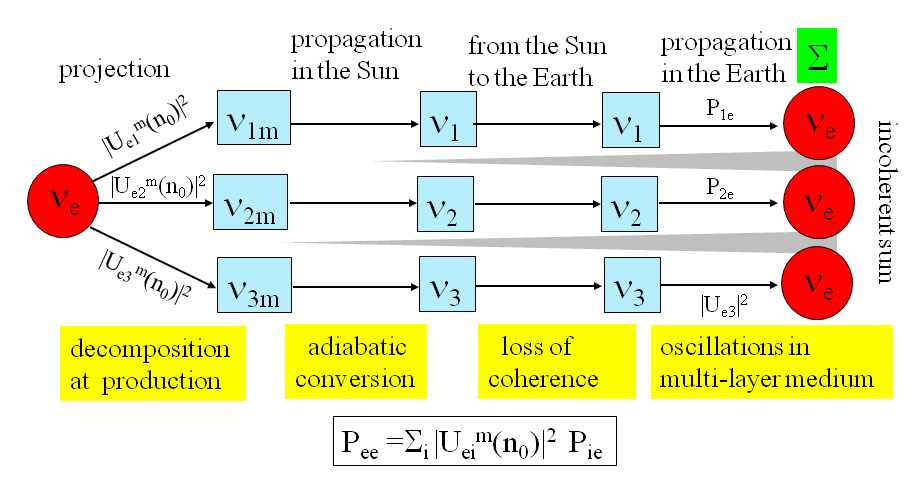}
\caption{Scheme of flavor transformations of solar neutrinos. 
The plot represents complete theory of the conversion. 
The shadowed triangles illustrate loss of coherence between the eigenstates. 
For the SNO energy range $\nu_e \approx \nu_{2m}$.} 
\label{fig:incoh}
\end{figure}

The SNO experiment has discovered 
effect of {\it the adiabatic flavor conversion} (the MSW effect).  
Oscillations (effect of the phase) are irrelevant. 
Evolution of the solar neutrinos can be considered  as independent 
(incoherent) propagation  of the produced eigenstates in matter (Fig.~\ref{fig:incoh}).  
Flavors of these eigenstates (described by mixing angle) change according to 
density change.  At high energies (SNO) the adiabatic conversion 
is close to the non-oscillatory transition which corresponds to production of 
single eigenstate.  Oscillations with small depth 
occur in the matter of the Earth. 


  


\section*{A year latter}

These remarks are written one year after the first submission of the paper. \\

1. Here is another attempt to explain things: In quantum mechanics in the two slits  
(equivalent to two eigenstates in our case) experiment one observes 
an interference picture (that is, oscillations).
In the case of one slit (one eigenstate) there is no interference 
(no-oscillations). The latter is close to what happens in the Sun with neutrinos 
observed by SNO.  Analogy to the SNO case would be one slit with
transparency depending on energy, if diffraction is neglected.
Should we speak about interference in the one slit case?\\

2.  In \cite{Bouchez:1986kb} the authors write ``as we shall see below, 
this (resonant oscillation - {\it A.S.}) is not exactly what happens in the sun...'', then referring to 
Messiah's description of the adiabatic approximation. \\

3. Not much has  changed during the year in spite of various discussions, publications  
in ``Science magazine'' \cite{ko} and even in Italian newspaper ``la Repubblica''. 
Considerations in this paper are persistently ignored,  still ``solar neutrinos oscillate'' 
just confirming that A. Messiah was right.


\section*{Acknowledgments} 

I would like to thank E. K. Akhmedov for valuable discussions and comments.

\section*{Appendix}

Here we provide some details for statements made in the text. \\

${\bf  A}$. The amplitudes of the wave packets  are proportional to the corresponding
mixing parameters in Eq. (\ref{eq:mutaumixing}).   In $\nu_\mu$ we have 
$g_2 = \cos\theta_{23} g$, $g_3 = \sin \theta_{23} g$, where $g$ is 
the shape factor without mixing. 
Since the admixture of $\nu_\tau$ in $\nu_2$ is given by
$\sin \theta_{23}$, the absolute value of amplitude
of   $\nu_\tau$ from $\nu_2$ equals
$(\sin \theta_{23}~\cos\theta_{23} g)$. Similarly,
since the admixture of $\nu_\tau$ in $\nu_3$ is  $\cos\theta_{23}$, 
we obtain the absolute value of amplitude
of   $\nu_\tau$ from $\nu_3$ 
$(\sin \theta_{23} ~\cos\theta_{23} g)$. The two amplitudes are equal.
Consequently, in the initial moment when the
oscillation phase is zero the two amplitudes cancel each other
completely.  So,  the probability to find $\nu_\tau$:
$P_{\mu \tau}(\phi_{osc}) = 0$. In the moment of time
when $\phi_{osc} = \pi$ tau parts from two WP interfere
constructively leading to the total amplitude
$2 \sin \theta_{23} \cos\theta_{23} = \sin 2\theta_{23}$.
Therefore the  total probability to find $\nu_\tau$ equals 
$P_{\mu \tau}(\phi_{osc} = \pi) = \sin^2 2\theta_{23}$.
This determines the depth of oscillations.
Dependence of the probability on the phase can be reconstructed
using the above results for $\phi_{osc} = 0$ and $\pi$:
$$
P_{\mu \tau}(\phi_{osc}) = \sin^2 2\theta_{23}
\frac{1}{2} (1 - \cos \phi_{osc}) = \sin^2 2\theta_{23}
\sin^2 \frac{\phi_{osc}}{2}.
$$
This leads to the result in Eq. (\ref{eq:surv}).

\vskip 3mm

${\bf B}.$ Eigenstates of the Hamiltonian and mixing:  
introduction of mixing in matter make sense 
once we can introduce these eigenstates. This is clearly possible 
for constant density. 
If density changes, the Hamiltonian depends on time, $H(t)$, 
so one can speak about the eigenstates of instantaneous Hamiltonian. 
The instantaneous eigenstates have sense if the mixing (density) changes 
slowly enough.  In this case the adiabatic regime is 
realized. Also one can compute corrections to the 
adiabatic results. If the density changes quickly, so that 
corrections are large and adiabatic perturbation theory 
is broken, introduction of these eigenstates has no sense. 
In the case  of constant or adiabatically changing density 
introduction of the eigenstates is useful for 
solution of the problem.

\vskip 3mm

${\bf C.}$ The adiabaticity condition is  the condition under which 
the transitions $\nu_{1m} \leftrightarrow \nu_{2m}$ can be neglected. 
It  has very simple expression in the resonance, 
where, in fact,  it is most important.  The width of the resonance
layer $\Delta r_R$  (the layer where  $\sin^2 2\theta_{12}^m > 1/2$
and the angle changes from $\pi/8$ to $3\pi/8$) should be larger
than the oscillation length in matter in the resonance $l_m^R$:
$$
\Delta r_R \sim \tan  2 \theta_{12}     
\left(\frac{\Delta dn }{n dx}\right)^{-1} > l_m^R,  
$$
$l_m^R = l_\nu/ \sin 2 \theta_{12}$. 
Under this condition  the system has enough time to  adjust itself 
to changes of external conditions (matter density).


\end{document}